\documentclass[12pt]{article}
\usepackage{amssymb,amsmath,epsfig}
\usepackage{cite}
 \allowdisplaybreaks

\begin{document}
\title{\bf New Definition of Complexity Factor in $f(R,T,R_{\mu\nu}T^{\mu\nu})$ Gravity}
\author{Z. Yousaf \thanks{zeeshan.math@pu.edu.pk}, M. Z. Bhatti \thanks{mzaeem.math@pu.edu.pk} and T. Naseer \thanks{tayyabnaseer48@yahoo.com}\\
Department of Mathematics, University of the Punjab,\\
Quaid-i-Azam Campus, Lahore-54590, Pakistan.}

\date{}

\maketitle
\begin{abstract}
This paper is devoted to present new definition of complexity factor
for static cylindrically symmetric matter configurations in
$f(R,T,R_{\mu\nu}T^{\mu\nu})$ gravity. For this purpose, we have
considered irrotational static cylindrical spacetime coupled with a
locally anisotropic relativistic fluid. After formulating
gravitational field and conservation equations, we have performed
orthogonal splitting of the Riemann curvature tensor. Unlike GR (for
spherical case) the one of the structure scalars $X_{TF}$, has been
identified to be a complexity factor. This factor contains effective
forms of the energy density, and anisotropic pressure components.
Few peculiar relations among complexity factor, Tolman mass and Weyl
scalar are also analyzed with the modified
$f(R,T,R_{\mu\nu}T^{\mu\nu})$ corrections.
\end{abstract}

\section{Introduction}

Albert Einstein proposed the general theory of relativity (GR) in 1915 and provided his well-known field equations in which matter and geometry were interlinked. According to his beliefs, our cosmos does not expand or contract with the passage of time, i.e., it is static in nature. After few years, in 1929, Edwin Hubble shown the accelerating expansion of our universe by performing an experiment on galaxies and found the red-shifted light coming from them which was an indication that these galaxies are moving away from each other. There have recently been some remarkable observations which pointed out the need of deeper understanding of our cosmic dynamics because of the extremely high amount of dark energy and dark matter, which is about $95$ percent of our whole universe. Several relativistic astrophysicists have established interests in studying different ways to describe the dark source elements of the universe \cite{copeland2006dynamics, nojiri2011unified,nojiri2017modified,capozziello2010beyond,capozziello2011extended,
de2010f,joyce2015beyond,cai2016f,bamba2015inflationary,yousaf2016influence,yousaf2016causes,shamir2019behavior}. In a recent era, a variety of different modified gravity theories which could be used to study the dynamical properties of expanding universe have been proposed.

The $f(R)$ theory, which is the forthright generalization of GR was
attained by replacing the Ricci scalar with its generic function in
an action function. In contrast to GR, Nojiri and Odintsov
\cite{nojiri2007introduction} asserted the stability of $f(R)$
theory by constructing various $f(R)$ models which are consistent
with specific solar system experiments to explore mysterious facets
of the universe. Bamba \emph{et al.} \cite{bamba2012dark} examined
the $\Lambda$CDM-like universe in the light of various models and
also investigated some properties of dark energy. In $f(R)$ gravity,
various researchers explored the role of different components which
produce irregularity in a self-gravitating system and analyzed
precise solutions to modified field equations by considering some
applications
\cite{sharif2014instability,bhatti2017gravitational,abbas2018complexity,doi:10.1142/S0217732319503334}.
In accordance with GR, Yousaf and Bhatti
\cite{yousaf2016electromagnetic} observed that $f(R)$ gravitational
models are highly acceptable to host more massive compact stars.

The $f(R,T)$ theory where matter lagrangian is an arbitrary function
of matter and geometry was then proposed by Harko \emph{et al.}
\cite{harko2011f} in which $T$ indicates the trace of
energy-momentum tensor. They analyzed the motion of some test
particles in this theory and found equations of motion for them via
variational principle. Baffou \emph{et al.}
\cite{baffou2015cosmological} found some cosmological solutions
which are in accordance with the observed data and the stability of
a particular model through some methods was also checked. Haghani
\emph{et al.} \cite{haghani2013further} proposed the more general
form of $f(R,T)$ gravity which they called as $f(R,T,Q)$ theory, in
which the matter lagrangian contains the strong dependence of
geometry and fluid. They also determined field equations in this
theory by considering Lagrange multiplier method. Odintsov and
S\'{a}ez-G\'{o}mez \cite{odintsov2013f} observed the role of strong
non-minimal connection of spacetime with geometry and discovered
that some extra terms of $f(R,T,Q)$ gravity permit our cosmic
expansion.

In $f(R,T,Q)$ gravity, some particular scalar and vector fields have
been considered by Ayuso \emph{et al.} \cite{ayuso2015consistency}
in which they studied the conditions for stability. Using some
peculiar solutions, Baffou \emph{et al.} \cite{baffou2016exploring}
investigated the stability of $f(R,T,Q)$ theory and concluded that
the extra curvature terms could be helpful to understand the early
phases of our cosmic evolution. They also discussed the stability of
some particular models in $f(R,T,Q)$ theory by obtaining their
solution through numerical techniques. Yousaf \emph{et al.}
\cite{yousaf2016stability,Yousaf2017,yousaf2017stability,yousaf2018dynamical}
studied the gravitational collapse in self-gravitating structures
and calculated their equations of motion in $f(R,T,Q)$ theory as
well as some relations with Weyl tensor. Bhatti \emph{et al.}
\cite{bhatti2019spherical,bhatti2019dissipative} studied the
gravitational collapse in $f(R,T,Q)$ theory and calculated few
constraints under which the systems would enter into the collapsing
phase.

A mixture of multiple components that can cause to generate
complications in any balanced self-gravitating structure in known as
complexity. Here, we have to describe the role of $f(R,T,Q)$ theory
on the existing outcomes of zero complexity measures in GR.
Different concepts of complexity can be seen in different fields of
science. Among those several definitions, L\'{o}pez-Ruiz \emph{et
al.}
\cite{lopez1995statistical,calbet2001tendency,catalan2002features}
offered this concept via entropy and information. Entropy is the
measurement of any system's disorderness while knowledge about a
system could be known as information. This concept has also been
then introduced through a term which is known as disequilibrium by
L\'{o}pez-Ruiz \emph{et al.} \cite{lopez1995statistical}.

In physics, one can illustrate the concept of complexity by
considering simplest systems (which have no complexity by
definition), i.e., the isolated ideal gas and perfect crystal. These
both models are extreme (but opposite) in nature from each other.
The former model is completely disordered in its nature as it is
made-up of molecules which are moving randomly, and hence it
provides maximal information due to the equal participation of all
molecules in this system. In a later model, the constituents which
make the system are arranged in an ordered form and thus it gives
minimal data set as the study of its small portion is enough to know
about its nature. Hence, it could be seen that there is maximum
disequilibrium in former case while zero in later case. In
astrophysics, the structural properties of self-gravitating
structures can also be analyzed by using the concept of complexity
factor. Usually, the components which make the system more complex
are pressure, heat flux and energy density etc. Without considering
the pressure component in the stress energy tensor, only the energy
density is not sufficient to study complexity of the system.

The nature of various physical features has usually been tested
using cylindrical structures at different scales. The gravitational
collapse, its radiation, spinning celestial objects and rotating
fluids are especially in astrophysics, which inspire to take
cylindrical symmetry into account. The Birkhoff's principle states
that there is a vacuum outside a spherical symmetric object and thus
spherical fluid collapse induces no gravitational waves. That is why
one switches to another basic symmetry, namely cylindrical geometry.
For cylindrically symmetrical star, Einstein and Rosen
\cite{einstein1937gravitational} found the solutions for
gravitational waves by assuming weak gravitational fields and also
claimed that in the Minkowski space, such problem reduce to regular
cylindrical waves. Regarding cylindrical symmetric propagation,
several astrophysical aspects have been explored. In the collapse of
cylindrical star, Herrera and Santos \cite{herrera2005cylindrical}
investigated the conditions for smooth matching of inner and outer
manifold and justified the radial pressure on the boundary not to be
zero. But later, Herrera \emph{et al.} \cite{herrera2006matching}
found some mistakes in calculations and claimed that the correction
of those mistakes will give zero radial pressure on the boundary in
a result. There has been some interesting results about the
evolution and stability of relativistic systems
\cite{olmo2019stellar,yousaf2019role,sahoo2017anisotropic,
mishra2018anisotropic,sahoo2017wormholes,bhatti2020stability,yousaf2020construction,yadav2020existence}.

Herrera \emph{et al.} \cite{herrera2005static} found the matching
conditions for static self-gravitating cylindrical object with the
Levi-Civita vacuum geometry as well as equations of motion and
observed their regularity. They also shown that an incompressible
fluid can be represented by a set of conformally flat solutions.
Sharif and Butt \cite{sharif2012butt} described the complxity factor
for the static cylindrical system in GR. The expansion free
condition has been investigated for anisotropic cylindrical fluid
distribution by Yousaf and Bhatti \cite{yousaf2016cavity}, which
produces vacuum cavity inside the fluid distribution. Recently,
Herrera \cite{herrera2017gibbs,herrera2020landauer} and Yousaf
\emph{et al.} \cite{yousaf2019tilted,yousaf2019non} described the
importance of different congruences of observers for describing the
very different physical phenomena of the same geometry of fluids.

The paper is listed as below. We propose some new physical
parameters and field equations for $f(R,T,Q)$ gravitational theory
in the next section. After this, we find four scalar functions from
the Riemann tensor and claim one of them as complexity factor in
Sec. 3. In Sec. 4, we design the condition for disappearing
complexity factor and also provide some exact solutions of field
equations in the background of $f(R,T,Q)$ theory. Finally, all these
results are concluded with the effects of modified corrections in
Sec. 5.

\section{Fundamental Equations for Static Symmetric Cylinder}

The model is now being developed as static cylindrically symmetric in combination with anisotropic matter. By calculating $f(R,T,Q)$ equations of motion, we investigate some structural properties of such systems. Through C-energy and Tolman formalisms, we will also be able to elaborate our results. In order to describe such static structures, we define some physical variables. In addition, the Darmois junction conditions are to be evaluated on hypersurface $\Sigma$.

\subsection{Field Equations in $f(R,T,Q)$ Gravity}

The action in $f(R,T,Q)$ theory becomes \cite{odintsov2013f,ayuso2015consistency,baffou2016exploring}
\begin{equation}\label{e1}
S=\frac{1}{2}\int \sqrt{-g}\left[f(R,T,Q)+L_{m}\right]d^{4}x,
\end{equation}
where the matter Lagrangian is represented by $L_{m}$. In this case, we define it as $L_{m}=-\mu$, where $\mu$ is the energy density of the fluid configuration.

The field equations yield from action \eqref{e1} by varying it with respect to $g_{\gamma\rho}$ as
\begin{align}\nonumber
\left(f_{R}\right.&\left.-L_{m}f_{Q}\right)G_{\gamma\rho}+\left[\frac{1}{2}Rf_{R}+\Box f_{R}+L_{m}f_{T}-\frac{1}{2}f
+\frac{1}{2}\nabla_{\alpha}\nabla_{\pi}(f_{Q}T^{\alpha\pi})\right]g_{\gamma\rho}\\\nonumber
&-\nabla_{\gamma}\nabla_{\rho}f_{R}+\frac{1}{2}\Box \left(f_{Q}T_{\gamma\rho}\right)-\nabla_{\alpha}\nabla_{(\gamma}[T_{\rho)}^{\alpha}f_{Q}]
+2f_{Q}R_{\alpha(\gamma}T_{\rho)}^{\alpha}-\left(8\pi G+f_{T}\right.\\\label{e2}
&+\left.\frac{1}{2}Rf_{Q}\right)T_{\gamma\rho}-2(f_{Q}R^{\alpha\pi}+f_{T}g^{\alpha\pi})\frac{\partial^2 L_{m}}{\partial g^{\gamma\rho}\partial g^{\alpha\pi}}=0,
\end{align}
where $\nabla_\rho,~g_{\alpha\rho}$ and $R_{\alpha\rho}$ represent the covariant derivative, metric tensor and Ricci tensor respectively. Also, $\Box\equiv g^{\lambda\sigma}\nabla_\lambda\nabla_\sigma$ and $2R_{\alpha(\gamma}T_{\rho)}^{\alpha}=R_{\alpha\gamma}T_{\rho}^{\alpha}+R_{\alpha\rho}T_{\gamma}^{\alpha}$, in which we use the definition of symmetric bracket. In addition, the subscripts $R,~T$ and $Q$ show that we take partial derivative of the terms with respect to their arguments. In GR, a distinct connection between $R$ and $T$ can be provided by the trace of stress-energy tensor. We consider the matter stress-energy tensor $T_{\gamma\rho}$ given in Eq.\eqref{e2} as
\begin{equation}\label{e3}
T_{\gamma\rho}=\mu v_{\gamma} v_{\rho}-Ph_{\gamma\rho}+\Pi_{\gamma\rho},
\end{equation}
where
\begin{equation}\label{e4}
\Pi_{\gamma\rho}=\Pi\left(l_{\gamma} l_{\rho}+\frac{1}{3}h_{\gamma\rho}\right);\quad P=\frac{P_{r}+2P_{\bot}}{3},
\end{equation}
\begin{equation}\label{e5}
\Pi=P_{r}-P_{\bot};\quad h_{\gamma\rho}=\delta_{\gamma\rho}-v_\gamma v_{\rho},
\end{equation}
where $l^{\rho}$ denotes the four vector in radial direction and $h_{\gamma\rho}$ is the projection tensor. As, we have assumed anisotropic pressure so $\Pi$ represents it, in which radial and tangential pressures are denoted by $P_{r}$ and $P_{\bot}$ respectively. Note that generally in cylindrical anisotropic spacetime, the pressure may exist in three different directions, i.e., $P_{r},~P_{\bot}$ and $P_{z}$, but it is important to mention here that the stress energy tensor \eqref{e3} is not the most general form. The modified field equations may be written in relation with Einstein field equations as
\begin{equation}\label{e6}
G_{\gamma\rho}=8\pi T_{\gamma\rho}^{(eff)},
\end{equation}
where $G_{\gamma\rho}$ represents Einstein tensor and $T_{\gamma\rho}^{(eff)}$ could be seen as the energy-momentum tensor in $f(R,T,Q)$ gravity. In this case, however from Eq.\eqref{e2}, we found as follows
\begin{align}\nonumber
&3\Box
f_R-T(f_T+1)+\frac{1}{2}\Box(f_QT)+R(f_R-\frac{T}{2}f_Q)+\nabla_\lambda\nabla_\rho(f_QT^{\lambda\rho})\\\nonumber
&+2R_{\lambda\rho}T^{\lambda\rho}f_Q+(Rf_Q+4f_T)\textit{L}_m-2f-2\frac{\partial^2\textit{L}_m}{\partial g^{\lambda\sigma}\partial
g^{\lambda\rho}}\left(f_Tg^{\lambda\rho}+f_QR^{\lambda\rho}\right).
\end{align}
If $Q=0$ is presumed in the above expression, the relativistic effects of $f(R,T)$ theory can be observed, while one can analyze these results in $f(R)$ theory by considering vacuum case as well. The study of some celestial structures are presented in \cite{odintsov2013f,ayuso2015consistency,baffou2016exploring}, in which they have done their derivation and physical indication with detailed analysis.

We consider that our geometry has a boundary surface $\Sigma$ which separates the inner and outer regions of cylindrical spacetimes. So interior to $\Sigma$, the geometry may be configured as
\begin{equation}\label{e7}
ds^2=-Y^2 dr^2-r^2(d\theta^2+\alpha^2 dz^2)+X^2 dt^2,
\end{equation}
where $X=X(r)$ and $Y=Y(r)$. We enforce the above coordinates to illustrate cylindrical symmetry, so the ranges are
\begin{equation}\nonumber
0\leq r, \quad 0\leq\theta\leq 2\pi, \quad -\infty< z< +\infty, \quad -\infty\leq t\leq +\infty.
\end{equation}

We may define the four vectors in the comoving frame that suit the above system as
\begin{equation}\label{e8}
v^\rho=(X^{-1},0,0,0),\quad l^\rho=(0,Y^{-1},0,0),
\end{equation}
which must satisfy some relations, $l^\rho v_{\rho}=0,~l^\rho l_{\rho}=-1$. Also, the four-acceleration can be defined as $a^{\rho}=v^{\rho}_{;\pi}v^{\pi}$, all of whose components disappear except $a^{1}$ as
\begin{equation}\label{e9}
a_{1}=-\frac{X'}{X}.
\end{equation}

For cylindrical spacetime \eqref{e7}, the field equations in $f(R,T,Q)$ theory becomes
\begin{align}\label{e10}
\frac{8\pi}{(f_{R}-L_{m}f_{Q})}\mu^{(eff)}&=\frac{1}{r^2Y^2}-\frac{2Y'}{rY^3},\\\label{e11}
\frac{8\pi}{(f_{R}-L_{m}f_{Q})}P_{r}^{(eff)}&=-\frac{2X'}{rXY^2}-\frac{1}{r^2Y^2},\\\label{e12}
\frac{1}{(f_{R}-L_{m}f_{Q})}P_{\bot}^{(eff)}&=-\frac{X''}{XY^2}+\frac{X'Y'}{XY^3}+\frac{Y'}{rY^3}-\frac{X'}{rXY^2},
\end{align}
where $\mu^{(eff)},~P_{r}^{(eff)}$ and $P_{\bot}^{(eff)}$ contains the material variables with the modified corrections of $f(R,T,Q)$ gravity. We have added their values in Appendix A. Here, the derivative with respect to r is shown by prime.

It must be emphasized that, unlike the theory of GR and $f(R)$, the divergence of effective stress-energy tensor in $f(R,T,Q)$ theory is fade away, resulting in a loss of all equivalence principles. Thus, this theory involves non-geodesic motion of the moving molecules as an extra force is acting on these particles in its gravitational region. So, one can cast its value as
\begin{align}\nonumber
\nabla^\rho T_{\rho\lambda}&=\frac{2}{Rf_Q+2f_T+1}\left[\nabla_\lambda(\textit{L}_mf_T)
+\nabla_\lambda(f_QR^{\pi\rho}T_{\pi\lambda})-\frac{1}{2}(f_Tg_{\pi\sigma}+f_QR_{\pi\sigma})\right.\\\label{e13}
&\times\left.\nabla_\lambda T^{\pi\sigma}-G_{\rho\lambda}\nabla^\rho(f_Q\textit{L}_m)\right].
\end{align}
The equation for hydrostatic equilibrium can be calculated in cylindrical structure as
\begin{equation}\label{e14}
\left(\frac{P_{r}^{(eff)}}{H}\right)'=\frac{-X'}{HX}\left(\mu^{(eff)}+P^{(eff)}_{r}\right)
+\frac{2}{rH}\left(P^{(eff)}_{\bot}-P^{(eff)}_{r}\right)+ZY^2,
\end{equation}
where $H=f_{R}-L_{m}f_{Q}$. The term $Z$ given in Appendix A contains the additional curvature terms of modified gravity. The above equation could be regarded as the most general form of Tolman-Opphenheimer-Volkoff equation having anisotropic fluid which may give better understanding to the possible structural changes in the static cylindrical system.

The value of $\frac{X'}{X}$ can be extracted from Eq.\eqref{e11} as
\begin{equation}\label{e15}
\frac{X'}{X}=\frac{4\alpha r}{\alpha r-8m}\left(-\frac{4\pi r}{H}P^{(eff)}_{r}+\frac{m}{\alpha r^2}-\frac{1}{8r}\right).
\end{equation}
By utilizing Eq.\eqref{e15} in Eq.\eqref{e14}, we have
\begin{align}\nonumber
\left(\frac{P_{r}^{(eff)}}{H}\right)'&=\frac{4\alpha r}{H(\alpha r-8m)}\left(\frac{4\pi r}{H}P^{(eff)}_{r}-\frac{m}{\alpha r^2}+\frac{1}{8r}\right)\left(\mu^{(eff)}+P^{(eff)}_{r}\right)\\\label{e16}
&+\frac{2}{Hr}\left(P^{(eff)}_{\bot}-P^{(eff)}_{r}\right)+ZY^2,
\end{align}
where $m$ can be obtained through C-energy formula \cite{thorne1965energy} for cylindrical system as
\begin{equation}\label{e17}
m(r)\equiv \tilde{E}=E\hat{l}=\frac{1}{8}\left(1-\frac{4}{Y^2}\right),
\end{equation}
that can be described with the help of Eq.\eqref{e10} as
\begin{equation}\label{e18}
m(r)=\frac{\alpha r}{8}-4\pi\alpha\int_{0}^{r} \tilde{r}^2\left(\frac{\mu^{(eff)}}{H}\right) d\tilde{r}.
\end{equation}

The geometrical structure outside the hypersurface $\Sigma$ is defined by taking the metric of the form
\begin{equation}\label{e19}
ds^2=\frac{2M(\nu)}{r}d\nu^2-2drd\nu+r^2(d\theta^2+\alpha^2 dz^2),
\end{equation}
where total mass of the corresponding object is denoted by $M(\nu)$. We have found Darmois junction conditions and smooth matching criterion of the exterior and interior manifolds over the boundary for $f(R,T,Q)$ gravity, which is provided by Yousaf \emph{et al.} \cite{Yousaf2017} (after following Senovilla \cite{senovilla2013junction}). We establish some constraints at boundary surface $r=r_{\Sigma}$ as
\begin{align}\label{e20}
E-M_=^{\Sigma}\frac{1}{8}; \quad [P_{r}]_=^{\Sigma}-D_{0},
\end{align}
where Appendix A includes the value of $D_{0}$.

\subsection{Curvature Tensors}

One of the most common curvature tensor, i.e., the Riemann tensor can be expressed in terms of the Ricci tensor $R_{\gamma\rho}$, Weyl tensor $C_{\alpha\gamma\beta\rho}$ and the Ricci scalar $R$ as
\begin{eqnarray}\nonumber
R^\rho_{\mu\nu\alpha}&=&C^\rho_{\mu\nu\alpha}+\frac{1}{2}R^\rho_{\nu}g_{\mu\alpha}-\frac{1}{2}R^\rho_{\alpha}g_{\mu\nu}
+\frac{1}{2}R_{\mu\alpha}\delta^\rho_{\nu}-\frac{1}{2}R_{\mu\nu}\delta^\rho_{\alpha}\\\label{e21}
&+&\frac{1}{6}R\left(g_{\mu\nu}\delta^\rho_{\alpha}-\delta^\rho_{\nu}g_{\mu\alpha}\right).
\end{eqnarray}
The electric and magnetic part of Weyl tensor can be defined as
\begin{equation}\label{e22}
E_{\gamma\rho}=C_{\gamma\alpha\rho\delta}v^{\alpha}v^{\delta}, \quad H_{\gamma\rho}=\frac{1}{2}\eta_{\gamma\pi\mu\nu}C^{\mu\nu}_{\rho\sigma}v^{\pi}v^{\sigma},
\end{equation}
In this case of static cylindrical structure, the magnetic part of Weyl tensor is zero. The Weyl tensor can also be expressed as
\begin{equation}\label{e23}
C_{\rho\pi\kappa\rho}=(g_{\lambda\pi\alpha\beta}g_{\kappa\rho\gamma\delta}-\eta_{\lambda\pi\alpha\beta}\eta_{\kappa\rho\gamma\delta})v^\alpha v^\gamma E^{\beta\delta},
\end{equation}
where $g_{\gamma\rho\alpha\beta}=g_{\gamma\alpha}g_{\rho\beta}-g_{\gamma\beta}g_{\rho\alpha}$, and $\eta_{\gamma\rho\alpha\beta}$ is known as the Levi-Civita tensor. We may also write $E_{\gamma\rho}$ as
\begin{equation}\label{e24}
E_{\gamma\rho}=E\left(l_{\gamma}l_{\rho}+\frac{1}{3}h_{\gamma\rho}\right),
\end{equation}
and its non-vanishing components are
\begin{equation}\label{e25}
  E_{11}=\frac{2Y^2}{3}E, \quad E_{22}=-\frac{r^2}{3}E, \quad E_{33}=\alpha^2E_{22},
\end{equation}
where
\begin{equation}\label{e26}
E=-\frac{1}{2XY^2}\left[X''-\frac{X'Y'}{Y}+\frac{XY'}{rY}-\frac{X'}{r}+\frac{X}{r^2}\right],
\end{equation}
which must satisfy the following conditions
\begin{equation}\label{e27}
E^\gamma_{\gamma}=0,\quad E_{\gamma\rho}=E_{(\gamma\rho)},\quad E_{\gamma\rho}v^\rho=0.
\end{equation}

\subsection{The Mass Function}

Here, in order to analyze some properties of cylindrical structure, we shall calculate few intriguing equations by following the C-energy \cite{thorne1965energy} and Tolman \cite{tolman1930use} formalism. We will then find a relation among the mass function and the Weyl tensor. One can write with the use of Eqs.\eqref{e6}, \eqref{e17}, \eqref{e21} and \eqref{e24} as
\begin{equation}\label{e28}
m=\frac{\alpha r}{8}-\frac{4\pi\alpha}{3H}r^3\left(\mu^{(eff)}-P^{(eff)}_{r}+P^{(eff)}_{\bot}\right)+\frac{\alpha r^3E}{3},
\end{equation}
from which we can calculate as
\begin{equation}\label{e29}
E=\frac{4\pi}{r^3}\int_{0}^{r} \tilde{r}^3\left(\frac{\mu^{(eff)}}{H}\right)' d\tilde{r}-\frac{4\pi}{H} \left(P^{(eff)}_{r}-P^{(eff)}_{\bot}\right).
\end{equation}
The above expression depict the properties of cylindrical structure, such as inhomogeneous energy density and pressure anisotropy in correspondence with the Weyl tensor under some modified corrections. After using Eq.\eqref{e29} in Eq.\eqref{e28}, we get
\begin{equation}\label{e30}
m(r)=\frac{\alpha r}{8}-\frac{4\pi\alpha}{3H}r^3\mu^{(eff)}+\frac{4\pi\alpha}{3}\int_{0}^{r} \tilde{r}^3 \left(\frac{\mu^{(eff)}}{H}\right)' d\tilde{r}.
\end{equation}
This equation relate the mass function with the energy density homogeneity. Using this formula, one can study the cylindrical self-gravitating structure and analyze the effects of modified correction terms on the consequent changes arise due to the energy density inhomogeneity.

For a static matter configuration, Tolman \cite{tolman1930use} proposed another energy formula which is given as
\begin{equation}\label{e31}
m_{T}=4\pi\alpha \int_{0}^{r_{\Sigma}} XY\tilde{r}^2(T_{0}^{0(eff)}-T_{1}^{1(eff)}-2T_{2}^{2(eff)})d\tilde{r}.
\end{equation}
For self-gravitating symmetric structures, Bhatti \emph{et al.} \cite{zaeem2019energy,bhatti2019tolman} determined the expressions for Tolman mass in $f(R)$ gravity with and without including the effects of electromagnetic field. Tolman introduced this formula in order to find the total energy of the system and it becomes within the cylindrical distribution of radius $r$ as
\begin{equation}\label{e32}
m_{T}=4\pi\alpha \int_{0}^{r} XY\tilde{r}^2(T_{0}^{0(eff)}-T_{1}^{1(eff)}-2T_{2}^{2(eff)})d\tilde{r}.
\end{equation}
By making use of Eqs.\eqref{e10}-\eqref{e12} in above equation, we get
\begin{equation}\label{e33}
m_{T}=-\frac{X'r^2}{Y},
\end{equation}
and after substituting the value of $X'$ from Eq.\eqref{e15}
\begin{equation}\label{e34}
m_{T}=XY\left(\frac{4\pi}{H}r^3P_{r}^{(eff)}-\frac{m}{\alpha}+\frac{r}{8}\right).
\end{equation}
This expression for $m_{T}$ could be called as the effective gravitational mass. Also, the gravitational acceleration $(a=-l^{\rho}a_{\rho})$ of a particle which is under investigation is followed in a static gravitational region (field), which is instantly at rest as
\begin{equation}\label{e35}
a=\frac{X'}{XY}=-\frac{m_{T}}{Xr^2}.
\end{equation}
One can express Eq.\eqref{e33} in a more suitable way as
\begin{eqnarray}\label{e36}
m_{T}&=&(m_{T})_{\Sigma}(\frac{r}{r_{\Sigma}})^3+r^3\int_{r}^{r_{\Sigma}}\frac{XY}{\tilde{r}}\left[\frac{4\pi}{H}\Pi^{(eff)}-E\right]d\tilde{r},
\end{eqnarray}
or by utilizing the value of $E$ from Eq.\eqref{e29}
\begin{equation}\label{e37}
m_{T}=(m_{T})_{\Sigma}(\frac{r}{r_{\Sigma}})^3+r^3\int_{r}^{r_{\Sigma}}\frac{XY}{\tilde{r}}\left[\frac{8\pi}{H}\Pi^{(eff)}
-\frac{4\pi}{r^3}\int_{0}^{r}\tilde{r^3}\left(\frac{\mu^{(eff)}}{H}\right)'d\tilde{r}\right]d\tilde{r}.
\end{equation}
For cylindrical symmetric static spacetime, the last expression for Tolman mass having modified corrections could be very useful in order to interpret the role of effective pressure anisotropy, the Weyl scalar $E$ and effective irregularity in the energy density. Thus, it relates the phenomenon of inhomogeneous energy density and local anisotropic pressure in $f(R,T,R_{\gamma\rho}T^{\gamma\rho})$ gravity via Tolman mass.

\section{The Orthogonal Decomposition of The Riemann Curvature Tensor}

Bel \cite{bel1961inductions} and Herrera \emph{et al.} \cite{herrera2004spherically} suggested the orthogonal decomposition of the Riemann tensor. The following three tensors can be found through the above-mentioned method, as
\begin{align}\label{e38}
Y_{\gamma\rho}&=R_{\gamma\alpha\rho\delta}v^\alpha v^\delta,\\\label{e39}
Z_{\gamma\rho}&=*R_{\gamma\alpha\rho\delta}v^\alpha v^\delta=\frac{1}{2}\eta_{\gamma\alpha\epsilon\pi}R^{\epsilon\pi}_{\rho\delta}v^\alpha v^\delta,\\\label{e40}
X_{\gamma\rho}&=*R^*_{\gamma\alpha\rho\delta}v^\alpha v^\delta=\frac{1}{2}\eta_{\gamma\alpha}^{\epsilon\pi}R^*_{\epsilon\pi\rho\delta}v^\alpha v^\delta,
\end{align}
where $\eta_{\gamma\alpha}^{\epsilon\pi}$ is the well-known Levi-Civita symbol whose value is $-1,~1$ and $0$ for negative, positive and no permutation respectively, while steric serves as the dual operation on the subsequent tensor. Another form of the Riemann tensor \eqref{e21} could be expressed in terms of preceding tensors (see \cite{gomez2007dynamical}), after making use of field equations as
\begin{equation}\label{e41}
R^{\gamma\rho}_{\quad\delta\beta}=C^{\gamma\rho}_{\quad\delta\beta}+16\pi T^{(eff)[\gamma}_{[\delta} \delta_{\beta]}^{\rho]}+8\pi T^{(eff)}\left(\frac{1}{3}\delta^{\gamma}_{\quad[\delta} \delta_{\beta]}^{\rho}-\delta^{[\gamma}_{\quad[\delta} \delta_{\beta]}^{\rho]}\right),
\end{equation}
and by using the value of $T^{(eff)}_{\gamma\rho}$, one can also express Eq.\eqref{e41} as
\begin{equation}\label{e42}
R^{\gamma\rho}_{\quad\delta\beta}=R^{\gamma\rho}_{(I)\delta\beta}+R^{\gamma\rho}_{(II)\delta\beta}+R^{\gamma\rho}_{(III)\delta\beta},
\end{equation}
where
\begin{eqnarray}\nonumber
R^{\rho\gamma}_{(I)\alpha\delta}&=&\frac{16\pi}{H}(f_{T}+\frac{1}{2}Rf_{Q}+1) \left[\mu v^{[\rho}v_{[\alpha} \delta_{\delta]}^{\gamma]}- Ph^{[\rho}_{\quad[\alpha} \delta_{\delta]}^{\gamma]}+ \Pi^{[\rho}_{\quad[\alpha} \delta_{\delta]}^{\gamma]}\right]\\\nonumber
&+&\frac{8\pi}{H}\left[(f_{T}+\frac{1}{2}Rf_{Q}+1)(\mu-3P)+4\left\{\frac{R}{2}\left(\frac{f}{R}-f_{R}\right)+\mu f_{T}\right.\right.\\\nonumber
&-&\left.\frac{1}{2}\nabla_{\mu}\nabla_{\beta}(f_{Q}T^{\mu\beta})\right\}-2f_{Q}R_{\mu\pi}T^{\mu\pi}-\frac{1}{2}\Box\{f_{Q}(\mu-3P)\}-3\Box f_{R}\\\nonumber
&+&\left.\nabla_{\mu}\nabla_{\pi}(f_{Q}T^{\mu\pi})+2g^{\pi\xi}(f_{Q}R^{\mu\beta}+f_{T}g^{\mu\beta})\frac{\partial^2L_{m}}{\partial g^{\pi\xi}\partial g^{\mu\beta}}\right]\\\label{e43}
&\times&\left(\frac{1}{3}\delta^{\rho}_{\quad[\alpha} \delta_{\delta]}^{\gamma}-\delta^{[\rho}_{\quad[\rho} \delta_{\delta]}^{\gamma]}\right),
\end{eqnarray}
\begin{eqnarray}\nonumber
R^{\rho\gamma}_{(II)\alpha\delta}&=&\frac{4\pi}{H}\left[2\left\{\frac{R}{2}\left(\frac{f}{R}-f_{R}\right)+\mu f_{T}-\frac{1}{2}\nabla_{\pi}\nabla_{\beta}(f_{Q}T^{\pi\beta})\right\}\right.\\\nonumber
&\times&\left(\delta^{\rho}_{\alpha} \delta_{\delta}^{\gamma}-\delta^{\rho}_{\delta} \delta_{\alpha}^{\gamma}\right)-\frac{1}{2}\Box\left\{f_{Q}\left(T^{\rho}_{\alpha} \delta_{\delta}^{\gamma}-T^{\gamma}_{\alpha} \delta_{\delta}^{\rho}-T^{\rho}_{\delta} \delta_{\alpha}^{\gamma}+T^{\gamma}_{\delta} \delta_{\alpha}^{\rho}\right)\right\}\\\nonumber
&+&2\Box f_{R}\left(\delta^{\rho}_{\delta} \delta_{\alpha}^{\gamma}-\delta^{\rho}_{\alpha} \delta_{\delta}^{\gamma}\right)+\left( \delta_{\delta}^{\gamma}\nabla^{\rho}\nabla_{\alpha}- \delta_{\alpha}^{\gamma}\nabla^{\rho}\nabla_{\delta}- \delta_{\delta}^{\rho}\nabla^{\gamma}\nabla_{\alpha}\right.\\\nonumber
&+&\left. \delta_{\alpha}^{\rho}\nabla^{\gamma}\nabla_{\delta}\right)f_{R}
-f_{Q}\left(R^{\rho}_{\pi}T^{\pi}_{\alpha} \delta_{\delta}^{\gamma}-R^{\gamma}_{\pi}T^{\pi}_{\alpha} \delta_{\delta}^{\rho}-R^{\rho}_{\pi}T^{\pi}_{\delta} \delta_{\alpha}^{\gamma}+R^{\gamma}_{\pi}T^{\pi}_{\delta} \delta_{\alpha}^{\rho}\right)\\\nonumber
&-&f_{Q}\left(R_{\pi\alpha}T^{\pi\rho} \delta_{\delta}^{\gamma}-R_{\pi\alpha}T^{\pi\gamma} \delta_{\delta}^{\rho}
-R_{\pi\delta}T^{\pi\rho} \delta_{\alpha}^{\gamma}+R_{\pi\delta}T^{\pi\gamma} \delta_{\alpha}^{\rho}\right)\\\nonumber
&+&\frac{1}{2}\nabla_{\pi}\nabla_{\delta}\left\{f_{Q}\left(T^{\gamma\pi} \delta_{\alpha}^{\rho}-T^{\rho\pi} \delta_{\alpha}^{\gamma}\right)\right\}+\frac{1}{2}\nabla_{\pi}\nabla^{\gamma}\left\{f_{Q}\left(T^{\pi}_{\delta} \delta_{\alpha}^{\rho}-T^{\pi}_{\alpha} \delta_{\delta}^{\rho}\right)\right\}\\\nonumber
&+&\frac{1}{2}\nabla_{\pi}\nabla_{\alpha}\left\{f_{Q}\left(T^{\rho\pi} \delta_{\delta}^{\gamma}-T^{\gamma\pi} \delta_{\delta}^{\rho}\right)\right\}+\frac{1}{2}\nabla_{\pi}\nabla^{\rho}\left\{f_{Q}\left(T^{\pi}_{\alpha} \delta_{\delta}^{\gamma}-T^{\pi}_{\delta} \delta_{\alpha}^{\gamma}\right)\right\}\\\nonumber
&+&2g^{\rho\gamma}(f_{Q}R^{\pi\beta}+f_{T}g^{\pi\beta})\left\{\delta_{\delta}^{\gamma}\frac{\partial^2L_{m}}{\partial g^{\gamma\alpha}\partial g^{\pi\beta}}-\delta_{\delta}^{\rho}\frac{\partial^2L_{m}}{\partial g^{\rho\alpha}\partial g^{\pi\beta}}\right.\\\label{e44}
&-&\left.\left.\delta_{\alpha}^{\gamma}\frac{\partial^2L_{m}}{\partial g^{\gamma\delta}\partial g^{\pi\beta}}+\delta_{\alpha}^{\rho}\frac{\partial^2L_{m}}{\partial g^{\rho\delta}\partial g^{\pi\beta}}\right\}\right],
\\\label{e45}
R^{\rho\gamma}_{(III)\alpha\delta}&=&4v^{[\rho}v_{[\alpha} E_{\delta]}^{\gamma]}-\epsilon^{\rho\gamma}_{\pi}\epsilon_{\alpha\delta\beta}E^{\pi\beta},
\end{eqnarray}
with
\begin{equation}\label{e46}
\epsilon_{\rho\alpha\beta}=u^\nu \eta_{\nu\rho\alpha\beta},\quad \epsilon_{\rho\alpha\nu}u^\nu=0.
\end{equation}
As, we have considered the static cylindrical symmetric case so the magnetic part of Weyl tensor is zero.\\
Now, we are able to find three effective tensors, i.e., $Y_{\rho\sigma}, Z_{\rho\sigma}$ and $X_{\rho\sigma}$ by taking into consideration Eq.\eqref{e42} in $f(R,T,Q)$ gravity as
\begin{align}\nonumber
Y_{\rho\sigma}&=E_{\rho\sigma}+\frac{1}{H}\left\{\frac{4\pi}{3}(\mu+3P)h_{\rho\sigma}+4\pi \Pi_{\rho\sigma}\right\}(f_{T}+\frac{1}{2}Rf_{Q}+1)-\frac{8\pi}{3H}\\\nonumber
&\times\left\{\frac{R}{2}\left(\frac{f}{R}-f_{R}\right)+\mu f_{T}-\frac{1}{2}\nabla_{\alpha}\nabla_{\pi}(f_{Q}T^{\alpha\pi})\right\}h_{\rho\sigma}+\frac{4\pi}{H}\left[-\frac{1}{2}\right.\\\nonumber
&\times\left\{\Box(f_{Q}T_{\rho\sigma})-v_{\sigma}v^{\delta}\Box(f_{Q}T_{\rho\delta})-v_{\rho}v_{\gamma}\Box(f_{Q}T^{\gamma}_{\sigma})
+g_{\rho\sigma}v_{\gamma}v^{\delta}\Box(f_{Q}T^{\gamma}_{\delta})\right\}\\\nonumber
&+\left(\nabla_{\rho}\nabla_{\sigma}f_{R}-v_{\sigma}v^{\delta}\nabla_{\rho}\nabla_{\delta}
f_{R}-v_{\rho}v_{\gamma}\nabla^{\gamma}\nabla_{\sigma}f_{R}
+g_{\rho\sigma}v_{\gamma}v^{\delta}\nabla^{\gamma}\nabla_{\delta}f_{R}\right)\\\nonumber
&+f_{Q}\left\{R_{\rho\alpha}(Ph^{\alpha}_{\sigma}-\Pi^{\alpha}_{\sigma})-R^{\gamma}_{\alpha}(\mu v^{\alpha}v_{\gamma}h_{\rho\sigma}+v_{\rho}v_{\gamma}Ph^{\alpha}_{\sigma}-v_{\rho}v_{\gamma}\Pi^{\alpha}_{\sigma})\right\}\\\nonumber
&+f_{Q}\left\{R_{\alpha\sigma}(Ph^{\alpha}_{\rho}-\Pi^{\alpha}_{\rho})-R_{\alpha\delta}(\mu v^{\alpha}v^{\delta}h_{\rho\sigma}+v_{\sigma}v^{\delta}Ph^{\alpha}_{\rho}-v_{\sigma}v^{\delta}\Pi^{\alpha}_{\rho})\right\}\\\nonumber
&+\frac{1}{2}\{\nabla_{\alpha}\nabla_{\rho}(f_{Q}T^{\alpha}_{\sigma})+\nabla_{\alpha}\nabla_{\sigma}(f_{Q}T^{\alpha}_{\rho})
+g_{\rho\sigma}v_{\gamma}v^{\delta}\nabla_{\alpha}\nabla^{\gamma}(f_{Q}T^{\alpha}_{\delta})+g_{\rho\sigma}v_{\gamma}v^{\delta}\\\nonumber
&\times\nabla_{\alpha}\nabla_{\delta}(f_{Q}T^{\alpha\gamma})-v_{\sigma}
v^{\delta}\nabla_{\alpha}\nabla_{\rho}(f_{Q}T^{\alpha}_{\delta})
-v_{\gamma}v_{\rho}\nabla_{\alpha}\nabla_{\sigma}(f_{Q}T^{\gamma\alpha})-v_{\gamma}v_{\rho}\nabla_{\alpha}\nabla^{\gamma}\\\nonumber
&\times\left.(f_{Q}T^{\alpha}_{\sigma})-v_{\sigma}v^{\delta}
\nabla_{\alpha}\nabla_{\delta}(f_{Q}T^{\alpha}_{\rho})\}+2h^{\epsilon}_{\rho}(f_{Q}R^{\alpha\pi}+f_{T}g^{\alpha\pi})\frac{\partial^2L_{m}}{\partial g^{\epsilon\sigma}\partial g^{\alpha\pi}}\right]+\frac{8\pi}{3H}\\\nonumber
&\times\left[\frac{1}{2}\Box\{f_{Q}(\mu-3P)\}+2f_{Q}R_{\alpha\epsilon}(\mu v^{\alpha}v^{\epsilon}-Ph^{\alpha\epsilon}+\Pi^{\alpha\epsilon})
-\nabla_{\alpha}\nabla_{\epsilon}(f_{Q}T^{\alpha\epsilon})\right.\\\label{e47}
&-\left.2g^{\epsilon\xi}(f_{Q}R^{\alpha\pi}+f_{T}g^{\alpha\pi})\frac{\partial^2L_{m}}{\partial g^{\epsilon\xi}\partial g^{\alpha\pi}}\right]h_{\rho\sigma},
\\\nonumber
Z_{\rho\sigma}&=\frac{4\pi}{H}\left[\frac{1}{2}v^{\alpha}\Box(f_{Q}T^{\xi}_{\alpha})-v^{\alpha}\nabla^{\xi}\nabla_{\alpha}f_{R}+f_{Q}\mu R_{\alpha}^{\xi}v^{\alpha}-f_{Q}PR_{\alpha}^{\xi}v^{\alpha}+\frac{1}{3}f_{Q}\Pi R^{\xi}_{\alpha}v^{\alpha}\right.\\\label{e48}
&-\left.\frac{1}{2}v^{\alpha}\nabla_{\mu}\nabla^{\xi}(f_{Q}T^{\mu}_{\alpha})
-\frac{1}{2}v^{\alpha}\nabla_{\mu}\nabla_{\alpha}(f_{Q}T^{\mu\xi})\right]\epsilon_{\xi\sigma\rho},
\end{align}
and
\begin{align}\nonumber
X_{\rho\sigma}&=-E_{\rho\sigma}+\frac{1}{H}\left(\frac{8\pi}{3}\mu h_{\rho\sigma}+4\pi\Pi_{\rho\sigma}\right)(f_{T}+\frac{1}{2}Rf_{Q}+1)
+\frac{4\pi}{H}\left[\left\{-\frac{1}{2}\Box(f_{Q}T^{\pi}_{\epsilon})\right.\right.\\\nonumber
&+\left.\nabla^{\pi}\nabla_{\epsilon}f_{R}+\frac{1}{2}\nabla_{\alpha}\nabla^{\pi}(f_{Q}T^{\alpha}_{\epsilon})
+\frac{1}{2}\nabla_{\alpha}\nabla_{\epsilon}(f_{Q}T^{\alpha\pi})\right\}\epsilon^{\epsilon\delta}_{\rho}\epsilon_{\pi\delta\sigma}
+f_{Q}\left(P-\frac{\Pi}{3}\right)\\\nonumber
&\times\left.(R^{\pi}_{\alpha}\epsilon^{\alpha\delta}_{\rho}\epsilon_{\pi\delta\sigma}+R_{\alpha\epsilon})
\epsilon^{\epsilon\delta}_{\rho}\epsilon^{\alpha}_{\delta\sigma}\right]+\frac{8\pi}{3H}\left[\left\{\frac{R}{2}\left(\frac{f}{R}-f_{R}\right)+\mu f_{T}-\frac{1}{2}\nabla_{\alpha}\nabla_{\nu}\right.\right.\\\nonumber
&\times\left.(f_{Q}T^{\alpha\nu})\right\}-\frac{1}{2}\Box\{f_{Q}(\mu-3P)\}+2Rf_{Q}\left(P-\frac{\Pi}{3}\right)
+\nabla_{\alpha}\nabla_{\pi}(f_{Q}T^{\alpha\pi})\\\label{e49}
&+\left.2g^{\pi\epsilon}(f_{Q}R^{\alpha\nu}+f_{T}g^{\alpha\nu})\frac{\partial^2L_{m}}{\partial g^{\pi\epsilon}\partial g^{\alpha\nu}}\right]h_{\rho\sigma}.
\end{align}

Further, the four structure scalars $X_{T},~Y_{T},~X_{TF}$ and $Y_{TF}$ can then be received from $X_{\rho\sigma}$ and $Y_{\rho\sigma}$ to study some properties of cylindrical symmetry, as
\begin{align}\label{e50}
X_{T}&=\frac{8\pi\mu}{H}(f_{T}+\frac{1}{2}Rf_{Q}+1)+\psi_{1}^{(D)},
\\\label{e51}
X_{TF}&=-E+\frac{4\pi\Pi}{H}(f_{T}+\frac{1}{2}Rf_{Q}+1),
\\\label{e52}
Y_{T}&=\frac{4\pi}{H}(\mu+3P_{r}-2\Pi)(f_{T}+\frac{1}{2}Rf_{Q}+1)+\psi_{2}^{(D)},
\\\label{e53}
Y_{TF}&=E+\frac{4\pi\Pi}{H}(f_{T}+\frac{1}{2}Rf_{Q}+1)+\psi_{3}^{(D)}.
\end{align}
Equations \eqref{e51} and Eq.\eqref{e53} can also be composed by
using Eq.\eqref{e29} as
\begin{equation}\label{e54}
X_{TF}=-\frac{4\pi}{r^3}\int_{0}^{r}\tilde{r}^3\left(\frac{\mu^{(eff)}}{H}\right)'d\tilde{r}+\frac{4\pi\Pi^{(eff)}}{H}
+\frac{4\pi\Pi}{H}(f_{T}+\frac{1}{2}Rf_{Q}+1),
\end{equation}
and
\begin{equation}\label{e55}
Y_{TF}=\frac{4\pi}{r^3}\int_{0}^{r}\tilde{r}^3\left(\frac{\mu^{(eff)}}{H}\right)'d\tilde{r}-\frac{4\pi\Pi^{(eff)}}{H}
+\frac{4\pi\Pi}{H}(f_{T}+\frac{1}{2}Rf_{Q}+1)+\psi_{3}^{(D)}.
\end{equation}
where $\psi_{3}^{(D)}=\frac{1}{l_{\rho}l_{\sigma}+\frac{1}{3}h_{\rho\sigma}}\psi_{\rho\sigma}^{(D)}$. The values of $\psi_{1}^{(D)}$, $\psi_{2}^{(D)}$ and $\psi_{\rho\sigma}^{(D)}$ are added in Appendix B.

The effective anisotropic pressure in $f(R,T,Q)$ gravity can be expressed after using Eq.\eqref{e51} and Eq.\eqref{e53} again as
\begin{equation}\label{e56}
  X_{TF}+Y_{TF}=\frac{8\pi\Pi}{H}(f_{T}+\frac{1}{2}Rf_{Q}+1)+\psi_{3}^{(D)}.
\end{equation}
We utilize Eq.\eqref{e54} in Eq.\eqref{e36} to illustrate physical meanings of $X_{TF}$ as
\begin{eqnarray}\nonumber
m_{T}&=&(m_{T})_{\Sigma}\left(\frac{r}{r_{\Sigma}}\right)^3+r^3\int_{r}^{r_{\Sigma}}\frac{XY}{\tilde{r}}\left[X_{TF}
+\frac{4\pi\Pi^{(eff)}}{H}\right.\\\label{e57}
&-&\left.\frac{4\pi\Pi}{H}(f_{T}+\frac{1}{2}Rf_{Q}+1)\right] d\tilde{r}.
\end{eqnarray}

By matching Eq.\eqref{e37} with Eq.\eqref{e57}, we can see that Tolman mass contains the effects of the local anisotropic pressure and inhomogeneous energy density of the fluid provided by $X_{TF}$ under some modified correction terms of $f(R,T,Q)$ theory. Furthermore, an alternative way to express the Tolman mass is as
\begin{eqnarray}\nonumber
m_{T}&=&\int_{0}^{r}XY\tilde{r}^2\left[Y_{T}-\frac{4\pi}{H}(\mu+3P_{r}-2\Pi)(f_{T}
+\frac{1}{2}Rf_{Q}+1)\right.\\\label{e58}
&+&\left.\frac{4\pi}{H}(\mu^{(eff)}+3P_{r}^{(eff)}-2\Pi^{(eff)})-\psi_{2}^{(D)}\right]d\tilde{r}.
\end{eqnarray}
This expression explicitly relates $Y_T$ to the matter variables
with modified corrections and Tolman mass function in $f(R,T,Q)$
theory. Herrera \emph{et al.} \cite{herrera2011meaning,Herrera2012}
and Yousaf \emph{et al.}
\cite{PhysRevD.95.024024,bhatti2017dynamical,bhatti2017evolution}
shown the involvement of $Y_T$ in the evolutionary equation for the
expansion scalar, commonly known as Raychaudhuri equation. Thus,
Eq.\eqref{e58} may allow $m_T$ to be used even in $f(R,T,Q)$ gravity
to describe the Raychaudhuri equation.

\section{Matter Configuration With Disappearing Complexity Factor}

It is understood that in any static or non-static structures, various components could cause to produce complexity. The scalar $X_{TF}$ contains effective inhomogeneous energy density and effective anisotropic pressure which are sources to produce complexity in our system. In $f(R,T,Q)$ theory, five unknowns $(\mu,~X,~Y,~P_{r},~P_{\bot})$ are found in the relevant equations of motion. We therefore need to additional conditions to carry on our work. For this, one condition can be achieved through the disappearance of complexity factor \eqref{e54}, which gives
\begin{equation}\label{e59}
\Pi=\frac{H}{2}\left[\frac{1}{r^3}\int_{0}^{r}\tilde{r}^3\left(\frac{\mu^{(eff)}}{H}\right)'d\tilde{r}
-\frac{\Pi^{(D)}}{H}-\frac{\Pi}{H}(f_{T}+\frac{1}{2}Rf_{Q})\right].
\end{equation}
Here, two of the different models of stellar objects are taken into account. Now, we discuss these subcases as follows:

\subsection{The Gokhroo and Mehra Ans$\ddot{a}$tz}

We follow the assumption proposed by Gokhroo and Mehra \cite{gokhroo1994anisotropic} to continue the structural analysis as
\begin{equation}\label{e60}
\frac{\mu^{(eff)}}{H}=\mu_{o}\left(1-\frac{Kr^2}{r_{\Sigma}^2}\right),
\end{equation}
We will get following equation after using the above equation in Eq.\eqref{e18} as
\begin{equation}\label{e61}
m(r)=\frac{\alpha r}{8}+\beta r^3\left(-\frac{1}{3}+\frac{Kr^2}{5r_{\Sigma}^{2}}\right),
\end{equation}
and combine it with \eqref{e17}
\begin{equation}\label{e62}
Y=\sqrt{\frac{\alpha}{2\beta r^2\left(\frac{1}{3}-\frac{Kr^2}{5r_{\Sigma}^{2}}\right)}},
\end{equation}
with $K\in(0,1)$ and $\beta=4\pi\alpha\mu_{o}$. Moreover, we receive from Eqs.\eqref{e11} and \eqref{e12}
\begin{equation}\label{e63}
\frac{8\pi}{H}\left[P_{r}^{(eff)}-P_{\bot}^{(eff)}\right]=\frac{1}{Y^2}\left[-\frac{X'}{rX}
-\frac{1}{r^2}+\frac{X''}{X}-\frac{X'Y'}{XY}-\frac{Y'}{rY}\right].
\end{equation}
The new variables may helpful to introduce as
\begin{equation}\label{e64}
X^{2}=e^{\int(2z(r)-2/r)dr},\quad Y^{-2}=y(r),
\end{equation}
from which we get after replacing Eq.\eqref{e64} in Eq.\eqref{e63}
\begin{equation}\label{e65}
y'+y\left[\frac{2z'}{z}+2z-\frac{6}{r}+\frac{4}{r^2z}\right]=\frac{16\pi}{HZ}\Pi^{(eff)},
\end{equation}
that seems to be in Ricatti's suggested form. After integrating this expression in $f(R,T,Q)$ theory, the line element \eqref{e7} appears in a new form as
\begin{eqnarray}\nonumber
ds^2&=&-\frac{z^2(r)e^{\int \left(\frac{4}{r^2z(r)}+2z(r)\right)dr}}{r^6\left[16\pi\int\left\{\frac{z(r)\Pi^{(eff)}e^{\int \left(\frac{4}{r^2z(r)}+2z(r)\right)dr}}{Hr^6}\right\} dr+C\right]}dr^2\\\label{e66}
&-&r^2(d\theta^2+\alpha^2 dz^2)+e^{\int(2z(r)-2/r)dr}dt^2,
\end{eqnarray}
where $C$ is a constant of integration. Also, one may compose the effective physical variables in terms of Eq.\eqref{e64} as
\begin{align}\label{e67}
&\frac{4\pi \mu^{(eff)}}{H}=\frac{1}{2r^2}-\frac{m'}{\alpha r^2},
\\\label{e68}
&\frac{4\pi P^{(eff)}_{r}}{H}=\left(\frac{1}{2}-\frac{4m}{\alpha r}\right)\left(\frac{1}{4r^2}-\frac{z}{2r}\right),
\\\label{e69}
&\frac{4\pi P^{(eff)}_{\bot}}{H}=\left(\frac{1}{8}-\frac{m}{\alpha r}\right)\left(\frac{3z}{2r}-z'-z^2
-\frac{1}{r^2}\right)+\frac{m'z}{2\alpha r}.
\end{align}

Equations \eqref{e67},~\eqref{e68} and \eqref{e69} might be useful in cylindrical systems to understand their certain supernatural and very fascinating features. In the case of GR, Di Prisco \emph{et al.} \cite{di2011expansion} computed corresponding results, while their modified form for the cases of cylindrical and shearfree spherical structures were found by Sharif and Yousaf \cite{sharif2012shearfree,sharif2012expansion}. In the context of Einstein-$\Lambda$ gravity, Yousaf \cite{yousaf2017spherical} and Bhatti \cite{bhatti2016shear} further produced the extension of these results. The obtained solutions \eqref{e67}-\eqref{e69} should meet the condition \eqref{e20} at the boundary to achieve their non-singular characteristics.

\subsection{The Polytropic Fluid With Zero Complexity Factor}

Here, we analyze the existence of $f(R,T,Q)$ corrections in relativistic polytropic fluid. We have already assumed the vanishing complexity factor condition, but still to handle with such system of equations, we must require condition \eqref{e59} to supplement with the polytropic equation of state.  Here, we are going to discuss two cases of polytropes individually and the first of them is of the form
\begin{equation}\label{e70}
P^{(eff)}_{r}=K[\mu^{(eff)}]^{\gamma}=K[\mu^{(eff)}]^{(1+1/n)}; \quad X_{TF}=0,
\end{equation}
with $n,~K$ and $\gamma$ are the polytropic index, polytropic constant and polytropic exponent respectively.

It is easy to solve a system of equation in which the equations are in dimensionless form. In this regard, we introduce some variables to write the dimensionless form of TOV equation \eqref{e16} and the mass function as
\begin{align}\label{e71}
\sigma&=P^{(eff)}_{rc}/\mu^{(eff)}_{c},\quad r=\xi/A,\quad A^2=4\pi\mu^{(eff)}_{c}/\sigma(n+1),
\\\label{e72}
\phi^n&=\mu^{(eff)}/\mu^{(eff)}_{c},\quad \nu(\xi)=m(r)A^3/(4\pi\mu^{(eff)}_{c}),
\end{align}
where $\mu^{(eff)}_c$ and $P^{(eff)}_{rc}$ show their values at the center. At the boundary $r=r_{\Sigma}$(or $\xi=\xi_{\Sigma}$), we have $\phi(\xi_{\Sigma})=0$. After putting Eqs.\eqref{e71} and \eqref{e72} in TOV equation, we have
\begin{eqnarray}\nonumber
&&\frac{\xi^2}{1+\sigma\phi}\left[\alpha-\frac{8\nu\sigma(n+1)}{\xi}\right]\frac{d\phi}{d\xi}+\frac{2\Pi^{(eff)}
\phi^{-n}\xi}{P^{(eff)}_{rc}(n+1)(1+\sigma\phi)}\left[\alpha-\frac{8\nu\sigma(n+1)}{\xi}\right]\\\nonumber
&&+4\left(\nu-\frac{\alpha\sigma\xi^3\phi^{n+1}}{H}\right)-\frac{\alpha\xi}{2\sigma(n+1)}=\frac{\xi^2\phi^{-n}}{AP_{rc}^{(eff)}(n+1)(1+\sigma\phi)}\\\label{e73}
&&\times\left(Y^2ZH+\frac{A}{H}\phi^{n+1}P_{rc}^{(eff)}\frac{dH}{d\xi}\right)\left[\alpha
-\frac{8\nu\sigma(n+1)}{\xi}\right].
\end{eqnarray}
Differentiating $\nu$ with respect to $\xi$ and combine it with Eq.\eqref{e67}, we get
\begin{equation}\label{e74}
\frac{d\nu}{d\xi}=\frac{\alpha A^2}{32\pi\mu_{c}^{(eff)}}-\frac{\alpha\xi^2\phi^n}{H}.
\end{equation}

It is to be noted that three unknowns $\nu,~\phi$ and $\Pi$ are found in above two ordinary differential equations (ODE's) \eqref{e73} and \eqref{e74}. As we want to get the unique solution of such system of equations, we need one more condition and thus we rewrite condition \eqref{e59} in dimensionless form
\begin{align}\nonumber
&\frac{6\Pi}{n\mu^{(eff)}_{c}}+\frac{2\xi}{n\mu^{(eff)}_{c}}\frac{d\Pi}{d\xi}=\phi^{n-1}\xi\frac{d\phi}{d\xi}
+\frac{2\Pi\xi}{Hn\mu^{(eff)}_{c}}\frac{dH}{d\xi}-\frac{3}{n\mu^{(eff)}_{c}}\left[\Pi(f_{T}+\frac{1}{2}Rf_{Q})\right.\\\label{e75}
&+\left.\Pi^{(D)}\right]+\frac{H\xi\phi^n}{n}\frac{d}{d\xi}\left(\frac{1}{H}\right)
-\frac{H\xi}{n\mu^{(eff)}_{c}}\left[\frac{d}{d\xi}\left(\frac{\Pi^{(D)}}{H}\right)+\frac{d}{d\xi}\left\{\frac{\Pi}{H}(f_{T}+\frac{1}{2}Rf_{Q})\right\}\right].
\end{align}
Finally, we have a set of three equations \eqref{e73},~\eqref{e74} and \eqref{e75} and hence they provide a unique solution by giving arbitrary values to the parameters $n$ and $\sigma$. Also, one can study the physical characteristics such as mass, density and pressure of particular stellar object by solving the overhead equations for the particular values of their parameters $n$ and $\sigma$.

Now, we are ready to look upon the second case of equation of state for polytropes,i.e., $P^{(eff)}_{r}=K[\mu_{b}^{(eff)}]^{\gamma}=K[\mu_{b}^{(eff)}]^{(1+1/n)}$, where $\mu_{b}$ represents the baryonic mass density of the fluid. Hence, one can rewrite Eqs.\eqref{e73} and \eqref{e75} in this case as
\begin{align}\nonumber
&\frac{\xi^2}{1+\sigma\phi_{b}}\left[\alpha-\frac{8\nu\sigma(n+1)}{\xi}\right]\frac{d\phi_{b}}{d\xi}+\frac{2\Pi^{(eff)}
\phi_{b}^{-n}\xi}{P^{(eff)}_{rc}(n+1)(1+\sigma\phi_{b})}\left[\alpha-\frac{8\nu\sigma(n+1)}{\xi}\right]\\\nonumber
&+4\left(\nu-\frac{\alpha\sigma\xi^3\phi_{b}^{n+1}}{H}\right)-\frac{\alpha\xi}{2\sigma(n+1)}=\frac{\xi^2\phi_{b}^{-n}}{AP_{rc}^{(eff)}(n+1)(1+\sigma\phi_{b})}\\\label{e76}
&\times\left(Y^2ZH+\frac{A}{H}\phi_{b}^{n+1}P_{rc}^{(eff)}\frac{dH}{d\xi}\right)\left[\alpha-\frac{8\nu\sigma(n+1)}{\xi}\right],
\end{align}
and
\begin{align}\nonumber
&\frac{6\Pi}{n\mu^{(eff)}_{bc}}+\frac{2\xi}{n\mu^{(eff)}_{bc}}\frac{d\Pi}{d\xi}=\phi_{b}^{n-1}\xi\left[1+K(n+1)\left(\mu_{bc}^{(eff)}\right)^{1/n}\phi_{b}\right]
\frac{d\phi_{b}}{d\xi}+\frac{2\Pi\xi}{Hn\mu^{(eff)}_{bc}}\frac{dH}{d\xi}\\\nonumber
&-\frac{3}{n\mu^{(eff)}_{bc}}\left[\Pi(f_{T}+\frac{1}{2}Rf_{Q})+\Pi^{(D)}\right]+\frac{H\xi\phi_{b}^n}{n}\frac{d}{d\xi}\left(\frac{1}{H}\right)
-\frac{H\xi}{n\mu^{(eff)}_{bc}}\left[\frac{d}{d\xi}\left(\frac{\Pi^{(D)}}{H}\right)\right.\\\label{e77}
&+\left.\frac{d}{d\xi}\left\{\frac{\Pi}{H}(f_{T}+\frac{1}{2}Rf_{Q})\right\}\right],
\end{align}
with $\phi_{b}^n=\mu_{b}^{(eff)}/\mu_{bc}^{(eff)}$.

\section{Conclusions}

Our main goal to do this work is to study the structure of
self-gravitating cylindrical object under the influence of
correction terms of $f(R,T,Q)$ gravity. We took a static
cylindrically symmetric metric, and it is then assumed to be
combined with anisotropic matter. We then found the relevant
equations of motion
 an equation for the hydrostatic equilibrium in the realm of $f(R,T,Q)$ gravity.
 With the help of C-energy and Tolman mass formalisms, we determined some useful
 relations between $m$ and $m_{T}$. We split the Riemann tensor orthogonally in a
 proper way which results in a set of five $f(R,T,Q)$ scalar functions. We then
 explored their effects on the evolution and sustenance of matter distribution
 in the static relativistic cylinders. Recently, Herrera \cite{herrera2018new}
 presented the new concept of complexity for static spherically symmetric structures.
 The key assumption was that the system which contains homogeneous energy density
 coupled with isotropic fluid distribution will have less complexity. Hence,
 in this case, the derived effective scalar factor $X_{TF}$ is considered as
 the complexity factor in $f(R,T,Q)$ gravity. Now, we will draw attention to
 some important changes in modified gravity.\\

(i) The effective scalar function $X_{TF}$ contains the energy density
inhomogeneity and local pressure anisotropy with modified corrections of
$f(R,T,Q)$ gravity, which gives rise to the more complex system.\\

(ii) The term $X_{TF}$ could be useful to compute the Tolman mass in
terms of effective inhomogeneous energy density as well as effective
anisotropic pressure, as it has contributions of extra curvature terms of $f(R,T,Q)$ gravity.\\

(iii) For the case of dynamical dissipative fluid configuration, the effects
of heat flux may also involve in the scalar $X_{TF}$ with modified corrections.\\

After calculating modified equations of motion and Darmois junction
conditions, we get $X_{TF}$ \eqref{e54} as a complexity factor by
splitting the Riemann curvature tensor. Moreover, through the
disappearing complexity factor condition \eqref{e59} (by assuming
$X_{TF}=0$), we discussed two different applications to physical
systems. Firstly, we took an assumption on the energy density
suggested by Gokhroo and Mehra in order to investigate properties of
stellar structures in modified framework. In second example, we
dealt with the polytropic equation of state and wrote TOV equation
as well as condition \eqref{e59} in dimensionless form. The solution
of such equations under some constraints could help us to get better
understanding of a system. All our findings will significantly
reduce to GR by taking the constraint $f(R,T,Q)=R$
\cite{herrera2018new}.

For our observed static cylindrical system, we have found a
complexity factor mentioned in Eq.\eqref{e54}. The first term of
right hand side of Eq.\eqref{e54} describes the effective form of
inhomogeneous energy density and the negative sign represents that
this term is trying to make the system less complex. The second term
in this equation represents the local anisotropy pressure along with
correction terms of $f(R,T,Q)$ theory. This terms has been found to
be one of the main causes to make the system complex. It is seen
that this term would add the complexity in the cylindrical system if
$P_{r}>P_{\bot}$, while if during evolution $P_{r}<P_{\bot}$, then
this factor will tend to decrease the measure of complexity in the
system. The third term contains usual anisotropic pressure along
with the extra curvature terms of modified gravity. Thus our new
definition of complexity factor contains the effects of energy
density, anisotropic pressure along with the the corrections of
$f(R,T,Q)$ theory, which is quite different definition than that
found in GR for the spherical system by Herrera
\cite{herrera2018new}.

\vspace{0.25cm}

\section*{Appendix A}

The effective physical variables which have been appeared in Eqs.\eqref{e10}-\eqref{e12} are
\begin{align}\nonumber
\mu^{(eff)}&=\mu\left[1+2f_{T}+f_{Q}\left(\frac{1}{2}R-\frac{3X'}{rXY^2}-\frac{3X''}{2XY^2}+\frac{3X'Y'}{2XY^3}\right)+f'_{Q}\left(\frac{1}{rY^2}\right.\right.\\\nonumber
&-\left.\left.\frac{Y'}{2Y^3}\right)+\frac{f''_{Q}}{2Y^2}\right]
+\mu'\left[f_{Q}\left(\frac{1}{rY^2}-\frac{Y'}{2Y^3}\right)+\frac{f'_{Q}}{Y^2}\right]
+\frac{\mu''f_{Q}}{2Y^2}+P_{r}\left[f_{Q}\right.\\\nonumber
&\times\left.\left(\frac{X''}{2XY^2}-\frac{1}{r^2Y^2}+\frac{Y'}{rY^3}-\frac{X'Y'}{2XY^3}\right)+f'_{Q}\left(\frac{Y'}{2Y^3}
-\frac{2}{rY^2}\right)-\frac{f''_{Q}}{2Y^2}\right]\\\nonumber
&+P'_{r}\left[f_{Q}\left(\frac{Y'}{2Y^3}-\frac{2}{rY^2}\right)
-\frac{f'_{Q}}{Y^2}\right]-\frac{P''_{r}f_{Q}}{2Y^2}+P_{\bot}\left[f_{Q}\left(\frac{X'}{rXY^2}+\frac{1}{r^2Y^2}\right.\right.\\\label{e78}
&-\left.\left.\frac{Y'}{rY^3}\right)+\frac{f'_{Q}}{rY^2}\right]+\frac{P'_{\bot}f_{Q}}{rY^2}+\frac{R}{2}\left(\frac{f}{R}-f_{R}\right)+f'_{R}\left(\frac{2}{rY^2}-\frac{Y'}{Y^3}\right)+\frac{f''_{R}}{Y^2},
\\\nonumber
P_{r}^{(eff)}&=\mu\left[-f_{T}+f_{Q}\left(\frac{X'}{rXY^2}+\frac{X''}{2XY^2}-\frac{X'Y'}{2XY^3}\right)-\frac{f'_{Q}X'}{2XY^2}\right]-\frac{\mu'X'f_{Q}}{2XY^2}\\\nonumber
&+P_{r}\left[1+f_{T}+f_{Q}\left(\frac{1}{2}R+\frac{2X'}{rXY^2}+\frac{1}{r^2Y^2}-\frac{3X''}{2XY^2}+\frac{3Y'}{rY^3}+\frac{3X'Y'}{2XY^3}\right)\right.\\\nonumber
&+\left.f'_{Q}\left(\frac{X'}{2XY^2}+\frac{1}{rY^2}\right)\right]+P'_{r}\left[f_{Q}\left(\frac{X'}{2XY^2}+\frac{1}{rY^2}\right)\right]
+P_{\bot}\left[f_{Q}\left(\frac{Y'}{rY^3}\right.\right.\\\nonumber
&-\left.\left.\frac{X'}{rXY^2}-\frac{1}{r^2Y^2}\right)+\frac{3f'_{Q}}{2rY^2}\right]+\frac{P'_{\bot}f_{Q}}{rY^2}-\frac{R}{2}\left(\frac{f}{R}-f_{R}\right)-f'_{R}\\\label{e79}
&\times\left(\frac{X'}{XY^2}+\frac{2}{rY^2}\right),
\\\nonumber
P_{\bot}^{(eff)}&=\mu\left[-f_{T}+f_{Q}\left(\frac{X'}{rXY^2}+\frac{X''}{2XY^2}-\frac{X'Y'}{2XY^3}\right)
+\frac{f'_{Q}X'}{2XY^2}\right]+\frac{\mu'X'f_{Q}}{2XY^2}\\\nonumber
&+P_{r}\left[f_{Q}\left(\frac{X'}{rXY^2}+\frac{X''}{2XY^2}-\frac{X'Y'}{2XY^3}\right)
+f'_{Q}\left(\frac{X'}{XY^2}+\frac{1}{rY^2}-\frac{Y'}{2Y^3}\right)\right.\\\nonumber
&+\left.\frac{f''_{Q}}{2Y^2}\right]+P'_{r}\left[f_{Q}\left(\frac{X'}{XY^2}+\frac{1}{rY^2}-\frac{Y'}{2Y^3}\right)+\frac{f'_{Q}}{Y^2}\right]
+\frac{P''_{r}f_{Q}}{2Y^2}+P_{\bot}\left[1+f_{T}\right.\\\nonumber
&+\left.\left.f_{Q}\left(\frac{1}{2}R+\frac{2Y'}{rY^3}-\frac{2X'}{rXY^2}-\frac{2}{r^2Y^2}\right)
+f'_{Q}\left(\frac{X'}{2XY^2}-\frac{Y'}{2Y^3}\right)+\frac{f''_{Q}}{2Y^2}\right]\right.\\\nonumber
&+P'_{\bot}\left[f_{Q}\left(\frac{X'}{2XY^2}-\frac{Y'}{2Y^3}\right)
+\frac{f'_{Q}}{Y^2}\right]+\frac{P''_{\bot}f_{Q}}{2Y^2}-\frac{R}{2}\left(\frac{f}{R}-f_{R}\right)+f'_{R}\\\label{e80}
&\times\left(\frac{Y'}{Y^3}-\frac{1}{rY^2}-\frac{X'}{XY^2}\right)-\frac{f''_{R}}{Y^2}.
\end{align}

The term $Z$ exists in Eq.\eqref{e14} because the nature of $f(R,T,Q)$ gravity is non-conserved as
\begin{align}\nonumber
Z &= \frac{2}{\left(2+Rf_{Q}+2f_{T}\right)}\left[\frac{f'_{Q}P_{r}}{Y^2}\left(\frac{X'Y'}{2XY}+\frac{Y'}{rY}+\frac{X'}{rX}
+\frac{1}{2r^2}\right)\frac{P_{r}}{2Y^2}\left\{f_{Q}\left(\frac{2Y''}{rY}\right.\right.\right.\\\nonumber
&+\frac{X''Y'}{XY}-\frac{9X'Y'^2}{XY^2}-\frac{6Y'^2}{rY^2}+\frac{X'Y''}{XY}-\frac{X'^2Y'}{X^2Y}-\frac{4Y'}{r^2Y}
+\frac{2X''}{rX}-\frac{2X'}{r^2X}-\frac{2X'^2}{rX^2}\\\nonumber
&-\left.\left.\frac{4X'Y'}{rXY}-\frac{2}{r^3}\right)+2f'_{T}Y^2\right\}+\frac{P'_{r}}{2Y^2}\left\{f_{Q}\left(\frac{X'Y'}{XY}
-\frac{X''}{X}+\frac{2Y'}{rY}\right)+f_{T}Y^2\right\}\\\nonumber
&-\frac{3\mu'f_{T}}{2}-\mu f'_{T}-\frac{\mu'f_{Q}}{XY^2}\left(\frac{X''}{2}-\frac{X'Y'}{2Y}+\frac{Y'}{r}\right)
-\frac{P'_{\bot}}{r^2Y^2}\left\{f_{Q}\left(\frac{rY'}{Y}-\frac{rX'}{X}\right.\right.\\\label{e81}
&-\left.\left.\left.1\right)-f_{T}r^2Y^2\right\}+\left(\frac{2X'}{rX}+\frac{1}{r^2}\right)\left(\mu'f_{Q}+\mu f'_{Q}\right)\right].
\end{align}

The quantity $D_0$ in Eq.\eqref{e20} can be stated as
\begin{eqnarray}\nonumber
D_{0}&=&\mu\left[-\tilde{f}_{T}+\tilde{f}_{Q}\left(\frac{X'}{rXY^2}+\frac{X''}{2XY^2}-\frac{X'Y'}{2XY^3}\right)\right]
-\frac{\mu'X'\tilde{f}_{Q}}{2XY^2}+P_{r}\left[\tilde{f}_{T}+\tilde{f}_{Q}\right.\\\nonumber
&\times&\left.\left(\frac{1}{2}R+\frac{2X'}{rXY^2}+\frac{1}{r^2Y^2}-\frac{3X''}{2XY^2}+\frac{3Y'}{rY^3}+\frac{3X'Y'}{2XY^3}\right)\right]
+P'_{r}\left[\tilde{f}_{Q}\left(\frac{1}{rY^2}\right.\right.\\\nonumber
&+&\left.\left.\frac{X'}{2XY^2}\right)\right]+P_{\bot}\left[\tilde{f}_{Q}\left(\frac{Y'}{rY^3}-\frac{X'}{rXY^2}-\frac{1}{r^2Y^2}\right)\right]
+\frac{P'_{\bot}\tilde{f}_{Q}}{rY^2}-\frac{R}{2}\\\label{e82}
&\times&\left(\frac{f}{R}-\tilde{f}_{R}\right).
\end{eqnarray}

\section*{Appendix B}

The structure scalars \eqref{e50}-\eqref{e53} contains modified corrections which are found as
\begin{align}\nonumber
\psi_{1}^{(D)}&=\frac{4\pi}{H}\left[\left\{h^{\gamma}_{\pi}\Box(f_{Q}T^{\pi}_{\gamma})
-2h^{\gamma}_{\pi}\nabla_{\gamma}\nabla^{\pi}f_{R}-h^{\gamma}_{\pi}\nabla^{\pi}\nabla_{\alpha}(f_{Q}T^{\alpha}_{\gamma})
-h^{\gamma}_{\pi}\nabla_{\gamma}\nabla_{\alpha}\right.\right.\\\nonumber
&\times\left.(f_{Q}T^{\alpha\pi})\right\}-\left.2f_{Q}(R^{\pi}_{\alpha}h^{\alpha}_{\pi}
+R_{\alpha\gamma}h^{\gamma\alpha})\left(P-\frac{\Pi}{3}\right)\right]+\frac{8\pi}{H}\left[\left\{\frac{R}{2}\left(\frac{f}{R}-f_{R}\right)\right.\right.\\\nonumber
&+\left.\mu f_{T}-\frac{1}{2}\nabla_{\alpha}\nabla_{\rho}(f_{Q}T^{\alpha\rho})\right\}
-\frac{1}{2}\Box\{f_{Q}(\mu-3P)\}+2Rf_{Q}\left(P-\frac{\Pi}{3}\right)\\\label{e83}
&+\left.\nabla_{\alpha}\nabla_{\pi}(f_{Q}T^{\alpha\pi})+2g^{\pi\gamma}(f_{Q}R^{\alpha\rho}+f_{T}g^{\alpha\rho})\frac{\partial^2L_{m}}{\partial g^{\pi\gamma}\partial g^{\alpha\rho}}\right],
\\\nonumber
\psi_{2}^{(D)}&=-\frac{8\pi}{H}\left\{\frac{R}{2}\left(\frac{f}{R}-f_{R}\right)+\mu f_{T}-\frac{1}{2}\nabla_{\alpha}\nabla_{\nu}(f_{Q}T^{\alpha\nu})\right\}+\frac{4\pi}{H}\left[-\frac{1}{2}\right.\\\nonumber
&\times\left.\left\{\Box(f_{Q}T)-v^{\pi}v^{\sigma}\Box(f_{Q}T_{\pi\sigma})-v^{\rho}v_{\gamma}\Box(f_{Q}T^{\gamma}_{\rho})
+4v_{\gamma}v^{\sigma}\Box(f_{Q}T^{\gamma}_{\sigma})\right\}\right.\\\nonumber
&+\left(\Box f_{R}-v^{\pi}v^{\sigma}\nabla_{\pi}\nabla_{\sigma}f_{R}-v^{\rho}v_{\gamma}\nabla^{\gamma}\nabla_{\rho}f_{R}
+4v_{\gamma}v^{\sigma}\nabla^{\gamma}\nabla_{\sigma}f_{R}\right)+f_{Q}\\\nonumber
&\times\left\{R^{\rho}_{\alpha}(Ph^{\alpha}_{\rho}-\Pi^{\alpha}_{\rho})-3R^{\gamma}_{\alpha}\alpha v^{\alpha}v_{\gamma}\right\}
+f_{Q}\left\{R^{\pi}_{\alpha}(Ph^{\alpha}_{\pi}-\Pi^{\alpha}_{\pi})-3R_{\alpha\sigma}\alpha v^{\alpha}v^{\sigma}\right\}\\\nonumber
&+\frac{1}{2}\{\nabla_{\alpha}\nabla_{\pi}(f_{Q}T^{\alpha\pi})+\nabla_{\alpha}\nabla_{\rho}(f_{Q}T^{\alpha\rho})
+4v_{\gamma}v^{\sigma}\nabla_{\alpha}\nabla^{\gamma}(f_{Q}T^{\alpha}_{\sigma})+4v_{\gamma}v^{\sigma}\\\nonumber
&\times\nabla_{\alpha}\nabla_{\sigma}(f_{Q}T^{\alpha\gamma})-v^{\pi}v^{\sigma}\nabla_{\alpha}\nabla_{\pi}(f_{Q}T^{\alpha}_{\sigma})
-v_{\gamma}v^{\rho}\nabla_{\alpha}\nabla_{\rho}(f_{Q}T^{\gamma\alpha})-v_{\gamma}v^{\rho}\\\nonumber
&\times\left.\nabla_{\alpha}\nabla^{\gamma}(f_{Q}T^{\alpha}_{\rho})-v^{\pi}v^{\sigma}\nabla_{\alpha}\nabla_{\sigma}(f_{Q}T^{\alpha}_{\pi})\}
+2h^{\epsilon\rho}(f_{Q}R^{\alpha\nu}+f_{T}g^{\alpha\nu})\frac{\partial^2L_{m}}{\partial g^{\epsilon\rho}\partial g^{\alpha\nu}}\right]\\\nonumber
&+\frac{8\pi}{H}\left[\frac{1}{2}\Box\{f_{Q}(\mu-3P)\}+2f_{Q}R_{\alpha\epsilon}T^{\alpha\epsilon}
-\nabla_{\alpha}\nabla_{\epsilon}(f_{Q}T^{\alpha\epsilon})-2g^{\epsilon\xi}\right.\\\label{e84}
&\times\left.(f_{Q}R^{\alpha\nu}+f_{T}g^{\alpha\nu})\frac{\partial^2L_{m}}{\partial g^{\epsilon\xi}\partial g^{\alpha\nu}}\right],
\\\nonumber
\psi_{\rho\sigma}^{(D)}&=-\frac{2\pi}{H}\left[h^{\xi}_{\rho}h^{\mu}_{\sigma}\Box(f_{Q}T_{\xi\mu})-\Box(f_{Q}T_{\rho\sigma})
-v_{\rho}v_{\sigma}v_{\gamma}v^{\delta}\Box(f_{Q}T^{\gamma}_{\delta})\right]+\frac{4\pi}{H}\\\nonumber
&\times\left[(h^{\xi}_{\rho}h^{\mu}_{\sigma}\nabla_{\mu}\nabla_{\xi}f_{R}-\nabla_{\rho}\nabla_{\sigma}f_{R}
-v_{\rho}v_{\sigma}v_{\gamma}v^{\delta}\nabla^{\gamma}\nabla_{\delta}f_{R})+2f_{Q}(h^{\xi}_{\rho}h^{\alpha}_{\sigma}R_{\xi\alpha}P\right.\\\nonumber
&-h^{\alpha}_{\sigma}R_{\rho\alpha}P-h^{\xi}_{\rho}R_{\xi\alpha}\Pi^{\alpha}_{\sigma}+R_{\rho\alpha}\Pi^{\alpha}_{\sigma})
+\frac{1}{2}\{h^{\xi}_{\rho}h^{\mu}_{\sigma}\nabla_{\alpha}\nabla_{\xi}(f_{Q}T^{\alpha}_{\mu})+h^{\xi}_{\rho}h^{\mu}_{\sigma}\nabla_{\alpha}\nabla_{\mu}\\\nonumber
&\times(f_{Q}T^{\alpha}_{\xi})-\nabla_{\alpha}\nabla_{\rho}(f_{Q}T^{\alpha}_{\sigma})-\nabla_{\alpha}\nabla_{\sigma}(f_{Q}T^{\alpha}_{\rho})
-v_{\rho}v_{\sigma}v_{\gamma}v^{\delta}\nabla_{\alpha}\nabla^{\gamma}(f_{Q}T^{\alpha}_{\delta})\\\nonumber
&-v_{\rho}v_{\sigma}v_{\gamma}v^{\delta}\nabla_{\alpha}\nabla_{\delta}(f_{Q}T^{\alpha\gamma})\}+2(f_{Q}R^{\alpha\nu}+f_{T}R^{\alpha\nu})h^{\epsilon}_{\rho}\\\label{e85}
&\times\left.\left\{h^{\mu}_{\sigma}\frac{\partial^2L_{m}}{\partial g^{\epsilon\mu}\partial g^{\alpha\nu}}-\frac{\partial^2L_{m}}{\partial g^{\epsilon\sigma}\partial g^{\alpha\nu}}\right\}\right].
\end{align}

\vspace{0.5cm}

{\bf Acknowledgments}

\vspace{0.25cm}

This work is supported by National Research Project for Universities
(NRPU), Higher Education Commission, Pakistan under the research
project No. 8754/Punjab/NRPU/R\&D/HEC/2017.


\vspace{0.3cm}

\begin{thebibliography}{10}

\bibitem{copeland2006dynamics}
E.~J. Copeland, M.~Sami, and S.~Tsujikawa {\em Int. J. Mod. Phys.
D}, vol.~15,
  p.~1753, 2006.

\bibitem{nojiri2011unified}
S.~Nojiri and S.~D. Odintsov {\em Phys. Rep.}, vol.~505, p.~59,
2011.

\bibitem{nojiri2017modified}
S.~Nojiri, S.~D. Odintsov, and V.~K. Oikonomou {\em Phys. Rep.},
vol.~692,
  p.~1, 2017.

\bibitem{capozziello2010beyond}
S.~Capozziello and V.~Faraoni, vol.~170.
\newblock Springer Science \& Business Media, 2010.

\bibitem{capozziello2011extended}
S.~Capozziello and M.~De~Laurentis {\em Phys. Rep.}, vol.~509,
no.~4,
  pp.~167--321, 2011.

\bibitem{de2010f}
A.~De~Felice and S.~Tsujikawa {\em Living Rev. Relativ.}, vol.~13,
p.~3, 2010.

\bibitem{joyce2015beyond}
A.~Joyce, B.~Jain, J.~Khoury, and M.~Trodden {\em Phys. Rep.},
vol.~568, p.~1,
  2015.

\bibitem{cai2016f}
Y.-F. Cai, S.~Capozziello, M.~De~Laurentis, and E.~N. Saridakis {\em
Rep. Prog.
  Phys.}, vol.~79, p.~106901, 2016.

\bibitem{bamba2015inflationary}
K.~Bamba and S.~D. Odintsov {\em Symmetry}, vol.~7, p.~220, 2015.

\bibitem{yousaf2016influence}
Z.~Yousaf, K.~Bamba, and M.~Z. Bhatti {\em Phys. Rev. D}, vol.~93,
no.~6,
  p.~064059, 2016.

\bibitem{yousaf2016causes}
Z.~Yousaf, K.~Bamba, and M.~Z. Bhatti {\em Phys. Rev. D}, vol.~93,
p.~124048,
  2016.

\bibitem{shamir2019behavior}
M.~F. Shamir and A.~Malik {\em Comm. Theor. Phys.}, vol.~71, p.~599,
2019.

\bibitem{nojiri2007introduction}
S.~Nojiri and S.~D. Odintsov {\em Int. J. Geom. Meth. Mod. Phys.},
vol.~4,
  no.~01, p.~115, 2007.

\bibitem{bamba2012dark}
K.~Bamba, S.~Capozziello, S.~Nojiri, and S.~D. Odintsov {\em
Astrophys. Space
  Sci.}, vol.~342, no.~1, p.~155, 2012.

\bibitem{sharif2014instability}
M.~Sharif and Z.~Yousaf {\em J. Cosmol. Astropart. Phys.},
vol.~2014, no.~06,
  p.~019, 2014.

\bibitem{bhatti2017gravitational}
M.~Z. Bhatti and Z.~Yousaf {\em Int. J. Mod. Phys. D}, vol.~26,
no.~06,
  p.~1750045, 2017.

\bibitem{abbas2018complexity}
G.~Abbas and H.~Nazar {\em Eur. Phys. J. C}, vol.~78, no.~6, p.~510,
2018.

\bibitem{doi:10.1142/S0217732319503334}
Z.~Yousaf {\em Mod. Phys. Lett. A}, vol.~34, no.~0, p.~1950333,
2020.

\bibitem{yousaf2016electromagnetic}
Z.~Yousaf and M.~Z. Bhatti {\em Mon. Not. Roy. Astron. Soc.},
vol.~458, no.~2,
  p.~1785, 2016.

\bibitem{harko2011f}
T.~Harko, F.~S.~N. Lobo, S.~Nojiri, and S.~D. Odintsov {\em Phys.
Rev. D},
  vol.~84, no.~2, p.~024020, 2011.

\bibitem{baffou2015cosmological}
E.~H. Baffou, A.~V. Kpadonou, M.~E. Rodrigues, M.~J.~S. Houndjo, and
J.~Tossa
  {\em Astrophys. Space Sci.}, vol.~356, no.~1, p.~173, 2015.

\bibitem{haghani2013further}
Z.~Haghani, T.~Harko, F.~S.~N. Lobo, H.~R. Sepangi, and S.~Shahidi
{\em Phys.
  Rev. D}, vol.~88, no.~4, p.~044023, 2013.

\bibitem{odintsov2013f}
S.~D. Odintsov and D.~S{\'a}ez-G{\'o}mez {\em Phys. Lett. B},
vol.~725, no.~4,
  pp.~437--444, 2013.

\bibitem{ayuso2015consistency}
I.~Ayuso, J.~B. Jim{\'e}nez, and {\'A}.~de~la Cruz-Dombriz {\em
Phys. Rev. D},
  vol.~91, no.~10, p.~104003, 2015.

\bibitem{baffou2016exploring}
E.~H. Baffou, M.~J.~S. Houndjo, and J.~Tosssa {\em Astrophys. Space
Sci.},
  vol.~361, no.~12, p.~376, 2016.

\bibitem{yousaf2018dynamical}Z. Yousaf, K. Bamba, M. Z. Bhatti and U. Farwa
{\em Eur. Phys. J. A}, \textbf{54}, 122 (2018).

\bibitem{yousaf2016stability}
Z.~Yousaf, M.~Z. Bhatti, and U.~Farwa {\em Mon. Not. Roy. Astron.
Soc.}, vol. 464, p. 4509, 2017.

\bibitem{Yousaf2017}
Z.~Yousaf, M.~Z. Bhatti, and U.~Farwa {\em Eur. Phys. J. C},
vol.~77, no.~6,
  p.~359, 2017.

\bibitem{yousaf2017stability}
Z.~Yousaf, M.~Z. Bhatti, and U.~Farwa {\em Class. Quantum Grav.},
vol.~34,
  no.~14, p.~145002, 2017.

\bibitem{bhatti2019spherical}
M.~Z. Bhatti, Z.~Yousaf, and M.~Nawaz {\em Int. J. Geom. Meth. Mod.
Phys.},
  vol.~17, p.~2050017, 2019.

\bibitem{bhatti2019dissipative}
M.~Z. Bhatti, K.~Bamba, Z.~Yousaf, and M.~Nawaz {\em J. Cosmol.
Astropart.
  Phys.}, vol.~09, p.~011, 2019.


\bibitem{lopez1995statistical}
R.~Lopez-Ruiz, H.~L. Mancini, and X.~Calbet {\em Phys. Lett. A},
vol.~209,
  no.~5-6, pp.~321--326, 1995.

\bibitem{calbet2001tendency}
X.~Calbet and R.~L{\'o}pez-Ruiz {\em Phys. Rev. E}, vol.~63, no.~6,
p.~066116,
  2001.

\bibitem{catalan2002features}
R.~G. Catal{\'a}n, J.~Garay, and R.~L{\'o}pez-Ruiz {\em Phys. Rev.
E}, vol.~66,
  no.~1, p.~011102, 2002.

\bibitem{einstein1937gravitational}
A.~Einstein and N.~Rosen {\em J. Franklin Inst.}, vol.~223, no.~1,
p.~43, 1937.

\bibitem{herrera2005cylindrical}
L.~Herrera and N.~O. Santos {\em Class. Quantum Grav.}, vol.~22,
no.~12,
  p.~2407, 2005.

\bibitem{herrera2006matching}
L.~Herrera, M.~A.~H. MacCallum, and N.~O. Santos {\em arXiv preprint
  gr-qc/0611147}, 2006.

\bibitem{olmo2019stellar} G. J. Olmo, D. Rubiera-Garcia, and A. Wojnar,
\emph{arXiv preprint arXiv:1912.05202}, 2019.

\bibitem{yousaf2019role}
Z.~Yousaf {\em Eur. Phys. J. Plus}, vol.~134, no.~5, p.~245, 2019.

\bibitem{sahoo2017anisotropic}
P.~K. Sahoo, P.~Sahoo, and B.~K. Bishi {\em Int. J. Geom. Meth. Mod.
Phys.},
  vol.~14, no.~06, p.~1750097, 2017.

\bibitem{sahoo2017wormholes}
P.~K. Sahoo, P.~H. R.~S. Moraes, and P.~Sahoo {\em Eur. Phys. J. C}
vol. 78, p. 46, 2018.

\bibitem{mishra2018anisotropic}
B.~Mishra, S.~Tarai, and S.~K. Tripathy {\em Mod. Phys. Lett. A},
vol.~33,
  p.~1850170, 2018.

\bibitem{bhatti2020stability}
M.~Z. Bhatti, Z.~Yousaf, and M.~Yousaf {\em Phys. Dark Universe},
vol.~28,
  p.~100501, 2020.

\bibitem{yousaf2020construction}
Z.~Yousaf {\em Phys. Dark Universe}, vol.~28, p.~100509, 2020.

\bibitem{yadav2020existence} A. K. Yadav, L. K. Sharma, B. K. Singh, and P. K. Sahoo,
{\em New Astr.}, vol. 78, p. 101382 , 2020.

\bibitem{herrera2005static}
L.~Herrera, G.~Le~Denmat, G.~Marcilhacy, and N.~O. Santos {\em Int.
J. Mod.
  Phys. D}, vol.~14, no.~03, p.~657, 2005.

\bibitem{sharif2012butt}
M.~Sharif and I.~I. Butt {\em Eur. Phys. J. C}, vol.~78, p.~850,
2018.

\bibitem{yousaf2016cavity} Z. Yousaf and M. Z. Bhatti, {\em Eur. Phys. J.
C}, vol. 76, p. 267, 2016.

\bibitem{herrera2017gibbs} L. Herrera, {\em Entropy}, vol. 19, p. 110, 2017.

\bibitem{herrera2020landauer} L. Herrera, {\em Entropy}, vol. 22, p. 340,
2020.

\bibitem{yousaf2019tilted} Z. Yousaf, M. Z. Bhatti, and S. Yaseen, {\em Eur. Phys. J.
Plus}, vol. 134, p. 487, 2019.

\bibitem{yousaf2019non} Z. Yousaf, M. Z. Bhatti, and M. F. Malik, {\em Eur. Phys. J.
Plus}, vol. 134, p. 470, 2019.


\bibitem{sharif2012expansion} M.~Sharif and
Z.~Yousaf {\em Can. J. Phys.}, vol.~90, p.~865, 2012.

\bibitem{sharif2014stability}
M.~Sharif and M.~Z. Bhatti {\em Phys. Lett. A}, vol.~378, no.~5,
p.~469,
  2014.

\bibitem{thorne1965energy}
K.~S. Thorne {\em Phys. Rev.}, vol.~138, no.~1B, p.~B251, 1965.

\bibitem{senovilla2013junction}
J.~M.~M. Senovilla {\em Phys. Rev. D}, vol.~88, p.~064015, 2013.

\bibitem{tolman1930use}
R.~C. Tolman {\em Phys. Rev.}, vol.~35, no.~8, p.~875, 1930.

\bibitem{zaeem2019energy}
M.~Z~Bhatti, Z.~Yousaf, and A.~Yousaf {\em Int. J. Geo. Meth. Mod.
Phys.},
  vol.~16, no.~1950041, 2019.

\bibitem{bhatti2019tolman}
M.~Z. Bhatti, Z.~Yousaf, and A.~Yousaf {\em Mod. Phys. Lett. A},
vol.~34,
  p.~1950012, 2019.

\bibitem{bel1961inductions}
L.~Bel in {\em Ann. Inst. Henri Poincar{\'e}}, vol.~17, p.~37, 1961.

\bibitem{herrera2004spherically}
L.~Herrera, A.~Di~Prisco, J.~Martin, J.~Ospino, N.~O. Santos, and
O.~Troconis
  {\em Phys. Rev. D}, vol.~69, no.~8, p.~084026, 2004.

\bibitem{gomez2007dynamical}
A.~G.-P. G{\'o}mez-Lobo {\em Class. Quantum Gravity}, vol.~25,
no.~1,
  p.~015006, 2007.

\bibitem{herrera2011meaning}
L.~Herrera {\em Int. J. Mod. Phys. D}, vol.~20, p.~2773, 2011.

\bibitem{Herrera2012}
L.~Herrera, A.~Di~Prisco, and J.~Ospino {\em Gen. Relativ. Gravit.},
vol.~44,
  no.~10, pp.~2645--2667, 2012.

\bibitem{PhysRevD.95.024024}
Z.~Yousaf, K.~Bamba, and M.~Z. Bhatti {\em Phys. Rev. D}, vol.~95,
p.~024024,
  2017.

\bibitem{bhatti2017dynamical}
M.~Z. Bhatti and Z.~Yousaf {\em Int. J. Mod. Phys. D}, vol.~26,
no.~04,
  p.~1750029, 2017.

\bibitem{bhatti2017evolution}
M.~Z. Bhatti, Z.~Yousaf, and M.~Ilyas {\em Eur. Phys. J. C},
vol.~77, no.~10,
  p.~690, 2017.

\bibitem{gokhroo1994anisotropic}
M.~K. Gokhroo and A.~L. Mehra {\em Gen. relativ. gravit.}, vol.~26,
p.~75,
  1994.

\bibitem{di2011expansion}
A.~Di~Prisco, L.~Herrera, J.~Ospino, N.~Santos, and
V.~Vi{\~n}a-Cervantes {\em
  Int. J. Mod. Phys. D}, vol.~20, p.~2351, 2011.

\bibitem{sharif2012shearfree}
M.~Sharif and Z.~Yousaf {\em Chin. Phys. Lett.}, vol.~29, p.~050403,
  2012.

\bibitem{yousaf2017spherical}
Z.~Yousaf {\em Eur. Phys. J. Plus}, vol.~132, p.~71, 2017.

\bibitem{bhatti2016shear}
M.~Z. Bhatti {\em Eur. Phys. J. Plus}, vol.~131, p.~428, 2016.

\bibitem{herrera2018new}
L.~Herrera {\em Phys. Rev. D}, vol.~97, p.~044010, 2018.

\end{thebibliography}
\end{document}